# Artificial Immune Privileged Sites as an Enhancement to Immuno-Computing Paradigm

Tejbanta Singh Chingtham[1]
[1] Department of Computer Science & Engineering, Sikkim Manipal Institute of Technology, Majitar, Sikkim, India.
Email: chingtham@gmail.com

G. Sahoo[2]
[2] Department of Information Technology, Birla Institute of Technology Mesra, Ranchi, Jharkhand, India
Email: gsahoo@bitmesra.ac.in

M.K.Ghose[3]
[3] Department of Computer Science & Engineering, Sikkim Manipal Institute of Technology, Majitar, Sikkim, India.
Email: mkghose@smuhmts.edu.in

*Abstract*— **The immune system is a highly parallel and distributed intelligent system which has learning, memory, and associative capabilities. Artificial Immune System is an evolutionary paradigm inspired by the biological aspects of the immune system of mammals**. **The immune system can inspire to form new algorithms learning from its course of action. The human immune system has motivated scientists and engineers for finding powerful information processing algorithms that has solved complex engineering problems. This work is the result of an attempt to explore a different perspective of the immune system namely the Immune Privileged Site (IPS) which has the ability to make an exception to different parts of the body by not triggering immune response to some of the foreign agent in these parts of the body. While the complete system is secured by an Immune System at certain times it may be required that the system allows certain activities which may be harmful to other system which is useful to it and learns over a period of time through the immune privilege model as done in case of Immune Privilege Sites in Natural Immune System.**

## I. Introduction

In recent years there has been considerable interest in exploring and exploiting the potential concept of biological systems for applications in computer science and engineering. This paper explores and attempts to model a system that is inspired by various aspects of the immune systems of mammals. Artificial immune system imitates the natural immune system that has sophisticated methodologies and capabilities to build computational algorithms that solves engineering problems efficiently. The main goal of the human immune system is to protect the internal components of the human body by fighting against the foreign elements such as the fungi, virus and bacteria [1]. However there are few parts in the body where the introduction of foreign agent initially brings in the attention of the immune system for attack but later on these foreign agents are not fought anymore and are considered as the part of the body. And these organs that allows such stimulation to take place is called the Immune Privilege Sites (IPS). This immune privileged site can be considered for implementation in computational algorithm just like other immune inspired computational algorithms. Considering the natural environment of computation data, there are same types of entity that can be counted as a threat or as a solution. The computation algorithm inspired by this mechanism provides the machine, the sense of differentiating between the helpful and harmless entity by itself. It provides a means, to the machine to decide the entity's kind by itself without any guidance or supervision whatsoever. The decision of the machine is actually governed by the set of rules which will vary from environment to environment.

## II. Immune Privilege Site

Immune privilege is a term used to describe certain sites in the body which are able to tolerate the introduction of antigen without eliciting an inflammatory immune response [13]. It is a phenomenon in which certain sites and certain tissues and organs fail to obey the rules of immunology. Foreign tissue grafts placed in immune privileged sites namely anterior chamber of the eye, brain, testis, pregnant uterus, enjoy extended, often indefinite, survival, but similar grafts placed at conventional body sites (skin, beneath kidney capsule) are rejected [13]. The grafts are normally recognized as foreign antigen by the body and attacked by the immune system. However in immune privileged sites, tissue grafts can survive for extended periods of time without any rejection taking place. Similarly, grafts prepared from immune privileged tissues (testis, cornea, brain) experience extended, often indefinite, survival when implanted at conventional body sites, whereas grafts prepared from conventional tissues (skin, heart, kidney) are rapidly rejected [2], [3]. There is an opinion that immune privilege is an evolutionary adaptation that provides vulnerable tissues incapable of regeneration with immune protection while avoiding loss of vital functions (such as vision, reproduction) [2], [4]. T-cells or T-lymphocytes belong to a group of white blood cells known as lymphocytes, and play a central role in cell-mediated immunity [13]. Antigens from immune privileged regions have been found to interact with the T-cells in an unusual way inducing tolerance as opposed to a destructive response [13]. Considering the Eye among the various sites demonstrating immune privileged site, one mechanism that operates to create immune privilege is systemic immune deviation that is generated in response to antigenic material placed in an immune privileged site. For the antigen presenting cells (APC) capture the antigen and migrate via the blood to the splenic marginal zone(spleen) and secrete chemokines (MIP-2) that attract NKT cells and other cells (marginal zone B cells, CD4+ and CD8+T cells). These multicellular clusters present eye-derived antigens to naive T cells that differentiate into regulatory, rather than effector cells [5],[13]. The regulator T cells act systemically to suppress the expression of Th1 and Th2 effector cells specific for the eye derived antigen. Despite the fact that Th1 and Th2 effectors are inhibited, CD8+ T cells are primed and B cells produce non-complement fixing antibodies to the eye derived antigen. Thus, immune deviation refers to the elicitation of humoral immunity in the absence of CD4+ T cell-mediated immunity [1]. In addition to the cells that are responsible for immune deviation, parenchymal cells of immune privilege sites and tissues (ocular pigment epithelium, corneal endothelium, testicular Sertolli cells, placental trophoblasts) express cell surface molecules (complement inhibitors, CD95 ligand) and secrete factors (TGF-β, neuropeptides) that suppress local expression of both innate and adaptive immune responses [5]. TGF-β has the property of immune suppressive and anti-inflammatory properties. TGF-β confers upon ocular APC the capacity to induce immune deviation. These causes an unusual behaviour of the Anterior Chamber called Anterior Chamber Associated Immune Deviation (ACAID). Similar reaction takes place in the various immune privileged sites of the body creating a tolerance for the tissue or cell by the immune system

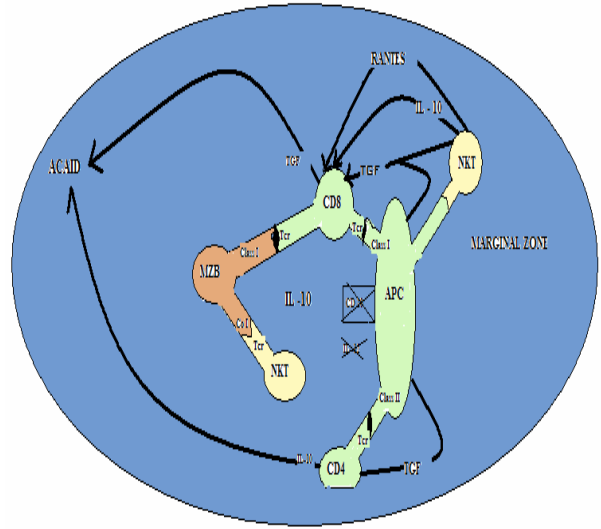

Fig 1. Spleen producing White Blood Cells

### III. IMMUNO-PRIVILEGE ALGORITHM

The immune system is highly complicated and appears to be precisely tuned to the problem of detecting and eliminating infections. It motivates the researchers to derive different kind of algorithm, drawing ideas from it. It is often noticed that artificial immune system implementation gives the most consistent result [9],[10]. Immune Privileged site are the sites that are made exception by the immune system. When an external agent is introduced in these region the immune system starts reacting against it but after some time it starts to consider it as a part of the body and stops reacting against it. On a similar line, we can create an algorithm that has been fed with a rule to consider the entity as a trusted entity by feeding itself information from its own initial action result on the entity. When an entity is initially introduced to a system, it reacts in a way as it is done to any foreign entity. The initial action on the entity is recorded and it is feed into a database with the its detail, if the action and resultant reaction performed on it fulfills the set of rules fed for the entity then it is set to be a trusted entity. Later when this same entity is introduced in the system environment it only checks if it is set as a trusted entity in the database. And hence the system doesn't attempt to discard or prevent this action of the entity. In this manner the system learns to differentiate between similar agents by its own experience of action and corresponding reactions. A generalized Immuno-Privileged Algorithm is described below:

**GENERALIZED ALGORITHM**

*input : $O_{observation}$ = Scan the Environment/Data.*
*output : D = Reaction of the system*
*memory : Trusted Entity/External Agent, set of rules to prepare the trust quotient(Tq)*

*begin*
  *repeat*
      *Check for any Entity/External Agent ($O_{observation}$)*
      *If $O_{observation}$ is not a trusted Entity/External Agent*
        *Then Do the normal task(D) on the Entity/External Agent, otherwise get disabled for the trusted Entity/External Agent*
        *If output done(D) on the Entity/External Agent matches the the rulesfor Tq*
        *Then Store the Entity/External Agent details in the trusted Entity/External Agent file database,*
        *Else proceed as normal*
        *End if*
      *End if*
  *until full job done*
*end*

*Environment: The region where the system is working.*

  *Agent/Entity: An Agent is anything that can be viewed as an object that is affecting the system to initiate some reaction other then its normal task of sensing the environment.*

### VI. ARTIFICIAL IMMUNE PRIVELEGE MODEL FOR ROBOTS

Artificial Immune system has already been implemented in robotics field [10]. This has shown very effective results on robots analytical behaviour. The immune system gives the bot the power to evolve. Moreover immune system supports one of the most consistent path finding algorithms [6], [9], [11]. The adaptation capacity of the algorithm gives it an edge over other known algorithm. Artificial Immune Privileged site is a relatively new unexplored concept of immuno-computing techniques. This brings another dimension of deciding a negative acknowledgement from its own actions. The robot normally acts with similar response to all external agent (viz, obstacles, intrusion etc) initially. After its initial action it checks if the external agent is one of the trusted agent from the predefined set of rules, if it complies to the set rules then it stores the details of the agent in it's memory so that the next time it encounters the same agent it does not react the same as it would for untrusted or unknown entity. In this experiment a simple surveillance bot is designed whose task is to guard after an object (red ball in this experiment). If this ball is removed by any external agent then it follows it till it is placed right in front of the robot. The same object that the surveillance bot is supposed to guard is also subjected to routine maintenance or cleaning. This bot initially follows the maintenance entity, which perform this task. On the repeating this routine, it detects some of the features of the maintenance entity and stores it in its database as trusted. In this experiment a blue colored block is shown to the robot to inform about the routine maintenance and hence the robot enters into an Immune Privilege Mode. After the maintenance the object is exposed to the robot thereby allowing the robot to enter into the normal mode by placing the object back at its initial position. Fig 2 through Fig 4 shows the implementation of the surveillance bot trying to follow the ball if it is removed from its original position. If it is placed at one position it does not move. Two boebots are introduced as a fail-safe mechanism wherein if any of the robots cannot detect due to its short range IR sensors at least one of the bot would be able to keep the ball engaged which will behave like the natural immune paradigm[7], [8].

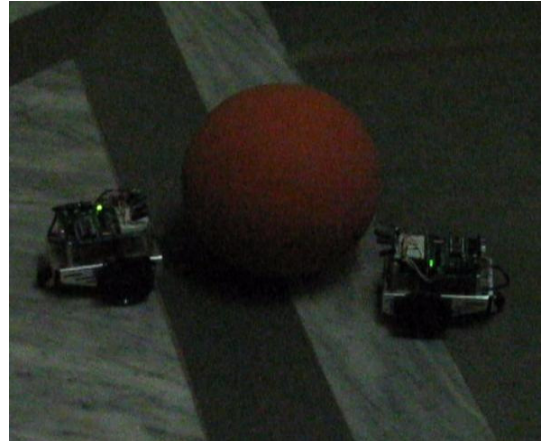
Fig 2 Two bots keeping an eye on the ball

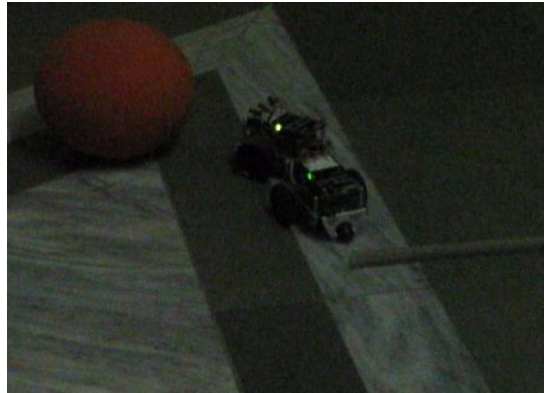
Fig 3 The ball is made to move using a stick and the surveillance bot start following the ball

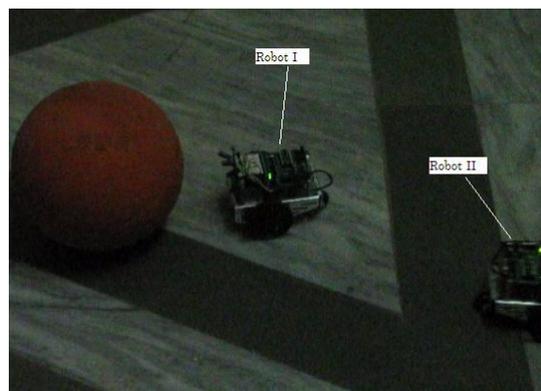
Fig 4 Robot II lost sight of the ball but bot I continue to keep the ball within its range

**Pseudo for the robotic implementation**

*Start*
*Store Initial Point*
*Flag = 0*
*loop:Store Position*
*Sense Front*
    *Sense Left*
      *Sense Right*
    *if(Front = Known Entity)*

*goto Disable*
*End if*
*If(Front=ObjDetect)*
*If(Flag = 1)*
    *Call Database*
*End if*
*goto Loop*
*End if*
*If(Left = ObjDetect & Front = ObjDetect)*
*If(Flag = 1)*
    *Call Database*
*End if*
*goto Loop*
*End if*
*if(Right=ObjDetect & Front= ObjDetect)*
*If(Flag = 1)*
    *Call Database*
*End if*
*goto Loop*
*End if*
*If (Left = ObjDetect & Front = ObjDetect & Right = ObjDetect)*
    *If(Flag = 1)*
        *Call Database*
    *End if*
*goto Loop*
*End if*
*If (Left = ObjDetect)*
    *Call Left*
    *Flag = 1*
*goto Loop*
    *End if*
    *If (Right = ObjDetect)*
        *CallRight*
        *Flag = 1*
        *goto Loop*
    *End if*
    *Flag = 1*
    *Call Forward*
    *goto Loop*
*Disable:Sense Front*
    *If(Front=KnownEntity)*
    *goto Disable*
*Else*
    *goto Loop*
    *End If*
*End*

*Start SubLeft*
*x = 0*
*y =0*
*Turn*
*Left*
*End Sub*

*Start SubRight*
*x = 0*
*y =0*
*TurnRight*
*End Sub*

*Start SubForward*
*x = 0*
*y = 0*
*if(x < Default_distance)*
    *Move Forward*
    *x = x+1*
 *End If*
 *End Sub*

*If(y < 360º)*
    *Turn Left*
    *y = y+5*
*End If*
*End Sub*

*x = 0*
*y = 0*
*End Sub*

*Start Sub Database*
*Flag = 0*
*intipos = intial point + (5/100 \*initial point)*
*intineg = intial point - (5/100 \*initial point)*
*for i = intineg to intipos*
*if(i = position)*
    *Known Entity = Entity*
*End if*
*End Sub*

## V. IMMUNE PRIVILEGED APPLICATION

The antivirus, anti malware and firewall are programmed to protect our computer/operating system from any kind of threat that can affect the function of the computer [11]. They are also used to protect the data from being stolen. These programs stops or removes the malicious programs from entering or running in the computer. Some of the programs that are normally not permitted by these protection software are viruses or strands of viruses, worms, keylogger, not trusted applications and many more of such malicious software. Some of these programs are deliberately installed in the computer for some reason by the user. Considering keylogger software, it is programmed to keep a log of all key strokes on the computer. It is used by many organizations to keep track of the work done by the employee with their knowledge on the organization's computer. This same program at times is not loaded by the user but by some external agent to intrude the user. This way the same software is sometimes loaded intentionally or sometimes loaded unintentionally. The antivirus or the operating system should have a way to recognize between the intentionally setup software and a software setup as a spy ware. This brings in the concept of artificial immune privileged computing that could be introduced to Antivirus or Anti-malware software. This concept provides the detecting software the capability to differentiate between unwanted malicious software and intentionally loaded software with malicious contents. All malicious software while being scanned by the antivirus/Anti-malware should have entered a unique captcha code along with the username and password by the user when popped as a threat by the antivirus. The captcha is used to differentiate between an external agent and the user, while the username & password is used to differentiate between different users working on the computer. Normally the antivirus detects the malicious software irrespective of its origin of installation. In this case the antivirus should check for the captcha for malicious software for the first time. It will ask the user to type in the captcha for the software if the user trusts the software. Once the user has typed in the captcha code then it allows the application to survive or does not faces deletion or moving to vault of the antivirus/anti-malware program and stores this software as a trusted application or in its positive database and no longer detects it as a threat in future. There by allowing the malicious looking program to enter into the immune privilege mode. Fig 5 shows a scenario wherein the Avira Antivirus detects the Keylogger application program which was deliberately installed on a computer being detected as a threat and prompting the action from the user [12]. Fig 6 shows the Keylogger program listed as malicious software in the threat database of Avira Antivirus software.

A Generic algorithm which maybe used to prevent such false alarm is described as follows

*Start*
*Loop:*
*If (system undergoing scan)*
    *Ip = Software/program being scanned*
    *Db = Database of unwanted program*
    *If (Ip = Db)*
        *Prompt as threat*
        *Prompt a Captcha*
        *If (input = prompted Captcha)*
            *Remove program from Db*
        *End If*
    *End If*
*End If*
*Goto loop*
*Stop*

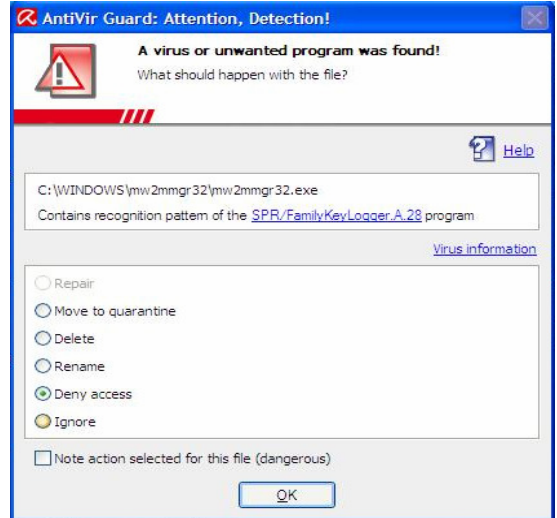

Fig 5 The antivirus detects the deliberately installed keylogger as an unwanted program

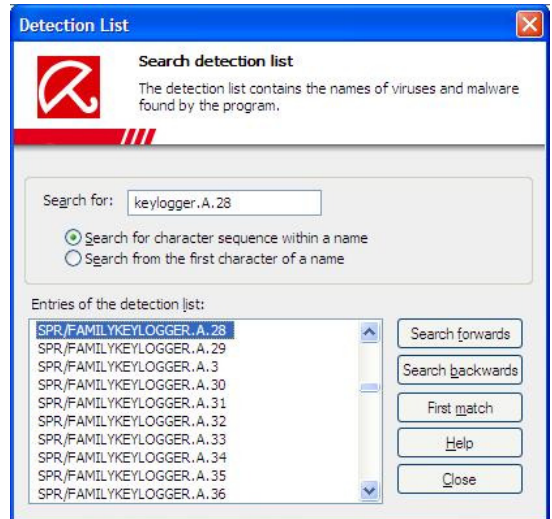

Fig 6 Keylogger listed as a threat in an Antivirus database

## VI. CONCLUSION

This research sheds light to a new field of Artificial immune system which gives the individual machine or system the capacity to learn and decide an external agent's intention to the system based on it's reaction on the agent.

The Artificial immune privilege site is a very new field that needs to be explored. The algorithm can further be improved to optimize and serve the need of different computational aspect.

## ACKNOWLEDGEMENT

This work is in part supported by AICTE-RPS grant vide Grant No 8023/BOR/RID/RPS-235/2008-09.

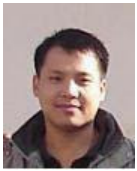

**Tejbanta Singh Chingtham** is associated with Sikkim Manipal University since August, 2000 and is working as an Associate Professor in the Department of Computer Science & Engineering. He received his B.Tech in Computer Science & Engineering from Bharathiar University, Coimbatore, India and M.Tech from Indian Institute of Technology Guwahati, India in Computer Science & Engineering and pursuing PhD from Birla Institute of Technology Mesra, Ranchi, India. He is a member of IEEE, IAENG and IACSIT. He has served as Technical Committee Members of Various Journals and Conferences. His main areas of Interest are Artificial Immune System, Natural Computing, Evolutionary Computation and Mobile Autonomous Robots, Navigation and Path Planning.

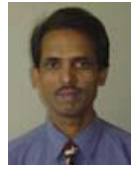

**G. Sahoo** received his M.Sc. degree in Mathematics from Utkal University in the year 1980 and Ph.D. degree in the area of Computational Mathematics from Indian Institute of Technology, Kharagpur in the year 1987. He has been associated with Birla Institute of Technology, Mesra, Ranchi,India since 1988 and is currently working as Professor and Head in the Department of Information Technology. His research interest includes Theoretical Computer Science, Parallel and Distributed Computing, Evolutionary Computing, Information Security, Image Processing and Pattern Recognition.

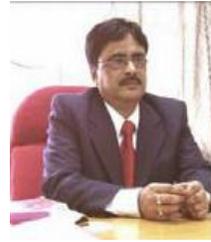

**M.K.Ghose** obtained his Ph.D. from Dibrugarh University, Assam, India in 1981. He is currently working as the Professor and Head of the Department of Computer Science & Engineering at Sikkim Manipal Institute of Technology, Mazitar, Sikkim, India. Dr. Ghose also served in the Indian Space Research Organization (ISRO) – during 1981 to 1994 at Vikram Sarabhai Space Centre, ISRO, Trivandrum in the areas of Mission simulation and Quality & Reliability Analysis of ISRO Launch vehicles and Satellite systems and during 1995 to 2006 at Regional Remote Sensing Service Centre, ISRO, IIT Campus, Kharagpur(WB), India in the areas of RS & GIS techniques for the natural resources management. Dr. Ghose has conducted a number of Seminars, Workshop and Training programmes and published around 95 technical papers in various national and international journals in addition to presentation/ publication of 125 research papers in international/ national conferences. He has supervised PhD desertations in areas like data mining, information security, geoinformatics and Robotics. He is a Life Member of Indian Association for Productivity, Quality & Reliability, Kolkata, National Institute of Quality & Reliability, Trivandrum, Society for R & D Managers of India, Trivandrum and Indian Remote Sensing Society, IIRS, Dehradun